\documentclass{rspublic}
\usepackage[numbers,comma,sort&compress]{natbib}
\usepackage{amssymb,bm,graphicx,rotating}
\newcommand{\beq}{\begin{equation}}
\newcommand{\eeq}{\end{equation}}
\newcommand{\dee}[1]{\,{\rm d}{#1}}

\begin{document} 
\title[Shock interaction and the IMF]{Shock interactions, turbulence and
the origin of the stellar mass spectrum} 
\author{Ralph E.~Pudritz \& N.K.-R.~Kevlahan} 
\affiliation{Origins Institute, McMaster University, Hamilton ON, Canada L8S 4M1} 
\label{firstpage} 
\maketitle

\begin{abstract}{Turbulence, interstellar medium, shocks, initial mass function.}
Supersonic turbulence is an essential element in understanding how structure
within interstellar gas is created and shaped.  In the context of star formation, many computational
studies show that the mass spectrum of density and velocity fluctuations within dense clouds,   
as well as the distribution of their angular momenta, trace their origin to the statistical and physical 
properties of gas that is lashed with shock waves.  In this article, we 
review the observations, simulations, and theories of how turbulent-like processes can account
for structures we see in molecular clouds.  We then compare traditional ideas of supersonic
turbulence with a simpler physical model involving the effects of multiple shock
waves and their interaction in the interstellar medium.  Planar intersecting shock
waves produce dense filaments, and generate vortex sheets that are 
essential to create the broad range of density and velocity structure in clouds.  As an example, 
the lower mass behaviour of the  stellar initial mass function can be traced to the tendency of a collection of shock waves to build-up a log-normal density distribution (or column density).   Vorticity -- which 
is essential to produce velocity structure over a very broad range of length scales in shocked clouds -- can also be
generated by the passage of curved shocks or intersecting planar shocks through such media.   
Two major additional physical forces affect the structure of star forming gas -- gravity and feedback processes from young stars.   Both of these can produce power-law tails at the high mass end of the IMF.  
\end{abstract}

\section{Overview: from gas to stars}

\begin{figure}
\[\includegraphics[width=\textwidth,clip=true]{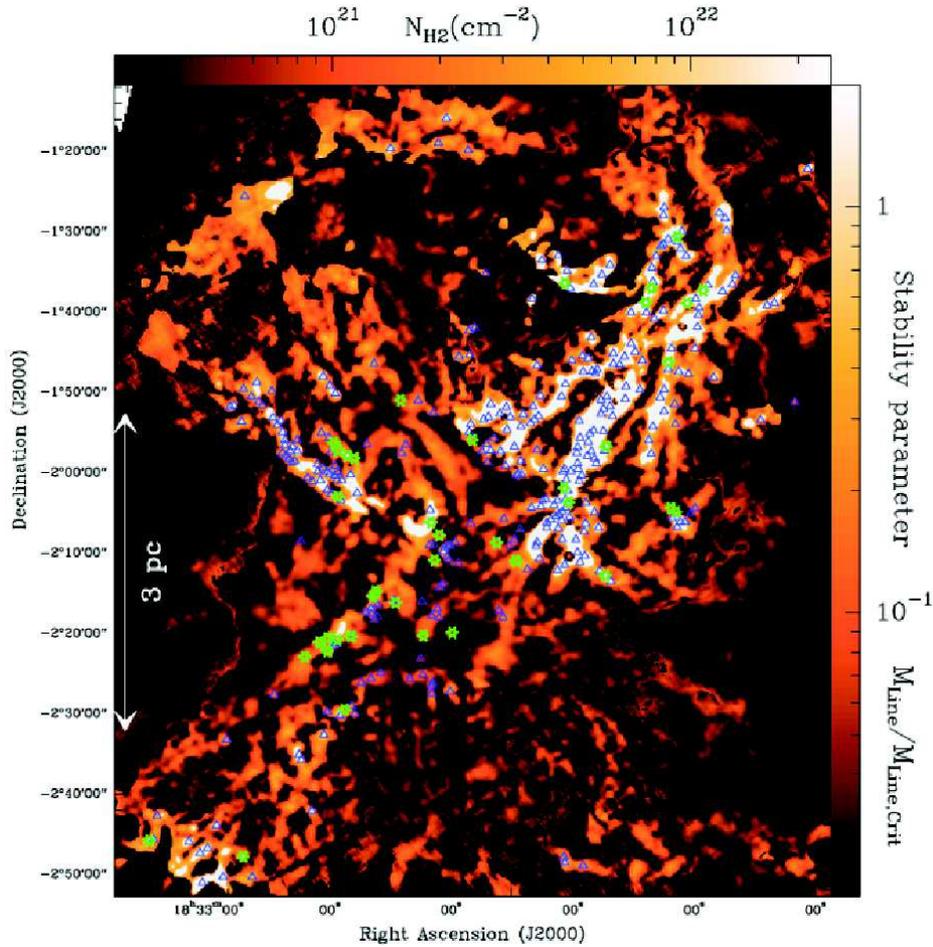}\]
\caption{Column density map of the Polaris Molecular Cloud, derived from Herschel Space Observatory data. Reproduced from~\cite{Andre/etal:2010}\label{fig:andre}.}
\end{figure}

The origin of stars in galaxies is directly coupled to the dynamics and structure of their interstellar gas.  What physical processes are responsible for converting gas that is, in its most diffuse form, non-self gravitating into a stellar mass spectrum (known as the initial mass function - IMF) with a well defined, possibly universal, simple form?    Both the diffuse (atomic and ionized) and molecular gas components of the interstellar medium (ISM) have highly supersonic velocity dispersions indicating that shocks are ubiquitous.  The propagation and collision of shock waves play an important role in the chain of events leading to star formation by compressing gas: pushing it into ever denser physical states that allows it to cool and fragment more efficiently.  This process also sweeps material into a network of dense filaments.  Both the compression and filamentary structure of over-dense regions help set up the conditions necessary for gravity to act.  Molecular clouds are sufficiently cold and dense that gravitational forces are a major factor in the evolution of the densest regions.   On small enough scales within such clouds (typically $\sim$1pc) in which star clusters are observed to form, gravity dominates and star formation occurs in the many dense gaseous ``cores'' which are evident in cluster forming regions.  Filamentary structure has always been apparent in observations of the interstellar medium \citep[e.g.][]{Schneider/Elmegreen:1979}, but the high resolution capabilities of the Herschel Space Observatory show that it is ubiquitous.  Most importantly, star formation takes place within these filaments~\cite{Menshchikov/etal:2010}.   Figure~\ref{fig:andre} shows the column density map of a molecular cloud as well as the positions of young stellar objects. The stars are forming along the gravitationally unstable parts of the filaments -- i.e. where the mass per unit length of the filament exceeds the critical value of $2c_s^2/G$~\cite{Andre/etal:2010} where $c_s$ is the 
sound speed of the gas and G is Newton's gravitational constant.

Several other physical processes also play important roles.  While we will not examine magnetic fields in detail in this paper, they are observed within both the diffuse and molecular components and within star-forming gas.  Magnetic fields in molecular gas have comparable, but somewhat, smaller energy densities than gravity \cite{Crutcher/etal:2010}.   The strength of magnetic fields is measured by the ratio of the gravitational to magnetic energy densities, known as the ``mass to flux'' ratio $\mu$.  Clouds and cores vary strongly in the value of $\mu$, but typically have values $\mu \le 5$, which has several important consequences.  One of these concerns the radial structure of density filaments, which are observed to have a much shallower dependence on radius than models of purely hydrodynamic, self-gravitating filaments \cite{Ostriker:1964}.   Good fits to the data are obtained for models of filaments that are wrapped with a helical magnetic field \cite{Fiege/Pudritz:2000}. Such structures have been detected by sub-millimetre polarimetry observations~\cite{Matthews/etal:2001}.  The formation of massive stars results in a copious outpouring of energy and momentum in the form of radiation fields and outflows which is pumped into the surrounding gas through the action of shocks.  Such ``feedback'' effects play a profound role in shutting down star formation and in energizing the large scale ISM.

Although several physical processes play a role in structuring molecular gas --- including turbulence, gravity, magnetic fields, and feedback due to stellar radiation and energetic protostellar outflows --- the spectrum of stellar masses that results  (known as the initial mass function, or IMF) is rather simple and appears to be universal.   The IMF is established at the scale at which star clusters are formed (a few parsecs).  It has been described by several different functional forms including a piece-wise power-law~\cite{Kroupa:2002}, a log-normal~\cite{Miller/Scalo:1979} and, more recently, as a log-normal with a high mass power-law tail~\cite{Chabrier:2003}.  All the studies of the IMF show that stars more massive than $0.5 M_{\odot}$ follow the famous Salpeter power-law form $dN \propto m^{-2.3}$. 

There have been many theoretical attempts to explain the particular form of the IMF.  For example, the log-normal distribution for the IMF at low masses could be due to gas supported by thermal pressure, while the Salpeter tail could be generated by the higher mass cores which would be supported by turbulent pressure
\citep[]{Hennebelle/Chabrier:2008}.  An alternative explanation is
that an initial log-normal distribution develops a power law tail as dense cores gradually accrete  \citep[]{Basu/Jones:2004}.  Elmegreen~\cite{Elmegreen:2002} remarked that if all gas above a certain density threshold in a log-normal distribution can form stars, then it is possible to recover the
well-known Schmidt law that governs the global star formation rate in galaxies \citep[see also][]{Wada/Norman:2007}. 

The IMF must ultimately derive from the structure of the molecular gas.  In this contribution we focus on the role that shocks and vorticity play in establishing the basic gas structure out of which the IMF ultimately arises.    While turbulence has often been invoked as a theoretical framework, we show that the spatial, density, and velocity structure of flows is much better understood in terms of the action of interacting shock waves.  This has several important ramifications for the general structure of the IMF.

\section{Understanding the observations of interstellar gas: supersonic turbulence vs. shocks}
Observations of chaotic, supersonic motions in interstellar gas have often been interpreted as arising from
``supersonic turbulence".  A basic property of turbulence is that it re-distributes energy and density over many length scales.   This process creates a strongly inhomogeneous flow which, in the context of molecular clouds, produces some regions that are much denser than others.  Sufficiently massive cores (greater than the local Jean's mass) will then collapse under their own gravity. This local collapse of dense cores begins the process of star formation.  In addition to transferring energy and density over a wide range of length scales (determined by the Reynolds number of the flow), turbulence also makes an initially uniform density very inhomogeneous in space.  This is because turbulence is extremely intermittent in space: the small scale regions (and hence energy dissipation) is concentrated on a small subset of space.  

The ISM, however, is more naturally pictured as a medium that is repeatedly lashed with shock waves that arise
from a wide range of energetic astrophysical processes.   Energy does not necessarily need to cascade from 
scale to scale: intersecting shocks and curved shocks can simultaneously excite energy over an enormous range
of length scales.  In addition, intersecting shocks and curved shocks efficiently generate vorticity.  We investigate
the observational constraints on these mechanisms below.  

\subsection{Spatial structure and density fluctuations}
On the largest galactic scales (several kpc) the overall structure of the shock-dominated, diffuse interstellar medium of the galaxy is seen in the neutral hydrogen (HI) gas observations such as the Canadian Galactic Plane survey~\cite{Taylor/etal:2003}.  These 21~cm maps of radio emission show that the diffuse ISM is highly filamentary, and abounds in bubbles, cavities, and channels carved out by radiation, winds, and supernova explosions from massive stars,.  Most stars form in star clusters, and it is the massive stars within them that heat and ionize the surrounding molecular and atomic gas through their intense radiation fields.  The feedback from radiation fields produces HII regions (ionized hydrogen) \cite[see e.g.][]{Peters/etal:2010}.  Towards the end of their short lives, massive stars produce stellar winds followed by supernova explosions.  Recent estimates \cite{Murray:2011} based on the properties of the largest GMCs in the galaxy, indicate that  the star formation efficiency in the galaxy is very low; in the range $\epsilon_{GMC} \simeq 0.002 - 0.2$ with estimated star formation rates per free-fall time of the molecular cloud being near unity.   The inference is that radiation pressure from massive stars in young, star clusters forming in these large GMCs very efficiently destroys most of the molecular gas that is collapsing and attempting to form stars.
Taken together,  feedback processes disperse significant amounts of molecular gas which probably accounts for the low star formation efficiency in galaxies.   A similar structure of the ISM is also observed in our satellite dwarf galaxy, the Large Magellanic Cloud (LMC) \citep[see][]{Elmegreen/etal:2001,Meixner/etal:2006}.  A bird's eye view of the relation between diffuse and molecular gas on the one hand, and the pattern of star formation and global density waves in a face-on disk galaxy on the other, is seen clearly in the HST image of M51. In this image the filamentary distribution of dense dust lanes that are associated with the spiral pattern are obviously associated with the occurrence of young star clusters.  

The prevalence of shock-driven processes extends down to the scales that characterize molecular clouds on tens of parsecs.  One of the best studied regions is the Orion molecular cloud.  The large scale structure of the Orion A and B clouds on 100~pc scales, mapped using an extinction technique \citep[e.g.][]{Cambresy:1999} is clearly filamentary.  The typical density of such gas is a few hundred cm$^{-3}$ which, though self-gravitating, is still too diffuse for star formation to occur.  We can follow the filamentary structure down to 10~pc scales wherein we see the famous integral filament, which was mapped at 850 microns by the JCMT sub-millimetre observatory \cite{Johnstone/Bally:2006}.    The dense, star forming gas in a molecular cloud occurs at 1~pc scales.  The typical density for such a region is $10^5 cm^{-3}$ and it is here that clusters of stars form.  Individual embedded infrared objects --- young stars in the process of formation --- are seen in dense fluctuations known as cores, within such cluster forming regions.  Long standing studies of the structure and physics of such cores \citep[e.g.][]{Jijina/etal:1999} show that they have length scales on the order of 0.04~pc and smaller.  Sufficiently low mass cores have subsonic velocity dispersion \cite{Pineda/etal:2010}.  The mass spectrum of cores has been analyzed for a number of different clouds \citep[e.g.][]{Motte/etal:2001,Johnstone/etal:2001,Nutter/Ward-Thompson:2007,Enoch/etal:2008} and has been found to be very similar in form to that of the IMF.  As an example, \cite{Alves/etal:2007} showed in their extinction map of the Pipe nebula that the core mass function could be overlaid on the IMF by shifting the curve down in mass by a factor of three.

The structure and scaling of turbulence intermittency is a major outstanding problem in turbulence theory~\cite{Alam/etal:2007}, although its existence is well-established.  Analysis of dust extinction maps~\cite{Kainulainen/etal:2009} have shown that the probability density function (PDF) of column density for star-forming molecular cloud complexes is well-described by a log-normal distribution, except at high densities where the PDF follows a power law with slopes in the range from roughly -1.5 to -4.2.   The density distribution can be converted to a mass distribution (called the initial mass function, IMF) if a length scale is associated to each density.  This can be done, for example, by assuming that the density fluctuations follow the same Kolmogorov scaling as the velocity, which seems to be the case.   The resulting PDF still has the form of a log-normal distribution with a power law tail at large masses.  The peak of this IMF is a mass characteristic of gravitational collapse \citep[e.g.][]{Hennebelle/Chabrier:2008,Padoan/Nordlund:2002}, while the power-law tail has a so-called Salpeter scaling exponent of about -1.35.

\subsection{Velocity structure and fluctuations}
Let us now look at the velocity structure of gas over these many scales.  The sparse gases and plasmas that fill interstellar space have velocity and mass density energy spectra characteristic of a neutral, incompressible fully-developed turbulent fluid~\cite{Elmegreen/Scalo:2004}.  For example, radio observations of the diffuse ISM have found that density fluctuations follow a so-called ``Big power law in the sky'': Kolmogorov-like scaling $k^{-5/3}$ of the energy spectrum extends over 11 orders of magnitude, from $10^6$m to $10^{17}$ m
\citep{Spangler:1999}!   Chepurnov and Lazarian~\cite{Chepurnov/Lazarian:2010} have recently performed a very careful analysis to show that the data are indeed consistent with a single Kolmogorov spectrum over the full range of length scales.  The ratio of the largest to smallest length scales increases like $Re^{3/4}$, where $Re$ is the Reynolds number.  Thus, these astrophysical flows must have an extremely high Reynolds number of about $10^{14}$ making them the most turbulent flows known.  Other fluctuations have power-law spectra with exponent $\alpha\in[-1.5,-2.6]$ \citep{Elmegreen/Scalo:2004}, also suggestive of (possibly two-dimensional) turbulence.   This is surprising because the interstellar medium has a very large mean free path, is conducting, has an embedded magnetic field, and is strongly compressible.  One would therefore expect the ISM to have very different dynamics from a simple Newtonian incompressible flow.  The origin of the turbulence-like properties of the ISM over at least 11 orders of magnitude remains one of the central mysteries of astrophysical fluid dynamics.

Some attempts have been made to explain the observed Kol\-mo\-go\-rov scaling in the context of a conducting compressible fluid.  A more comprehensive approach to incompressible MHD turbulence has
emphasized the anisotropy that is stamped on the turbulence by the magnetic 
fields  \cite{Sridhar/Goldreich:1994}.  This 
gives an energy spectrum $k_{\perp}^{-5/3}$ in directions
perpendicular to the mean magnetic field.  However, the theory is not
rigorous and it does not apply to the solar wind (which also exhibits a Kol\-mo\-go\-rov scaling).  Furthermore, other weak turbulence calculations find quite different scalings: $k_{\perp}^{-2}$ or $k_{\perp}^{-3/2}$. Stationary constant flux solutions exhibit exponents anywhere in the range from $-1$ to $-3$ depending on the asymmetry of the
forcing~\citep{Galtier/etal:2002}.  In addition, MHD turbulence is damped very quickly (in about one eddy turnover time).  Therefore any strongly turbulent flow would need a constant source of energy, or would be unlikely to reach a stationary statistical state (characteristic of so-called fully developed turbulence).  In summary, there is still no convincing explanation of why the ISM density and velocity fluctuations should follow a single Kolmogorov inertial range scale over such a wide range of scales.  We also note that recent high-resolution simulations of strongly magnetized MHD turbulence find a $k_\perp^{-3/2} $ scaling, rather than  Kolmogorov $k_\perp^{-5/3}$ scaling.   As suggested by Boldyrev et al~\citep{Boldyrev/etal:2009}, this discrepancy could be due to scale-dependent dynamic alignment of the velocity and magnetic field vectors. 

The Kolmogorov turbulence interpretation of the astrophysical data is also inconsistent with high resolution simulations.  As Johnsen et al.~\cite{Johnsen/etal:2010} point out in their comprehensive review of high-resolution methods for numerical simulations of compressible turbulence with shock waves, compressible turbulence is characterized by weak shock waves (eddy shocklets) that develop spontaneously from the turbulent fluctuations.  Shocklets were first observed in the numerical simulations performed by Kida \& Orszag in 1990~\cite{Kida/Orszag:1990}.  These shocklets develop when the turbulence Mach number (based on the r.m.s. velocity) $M_t>0.3$.  Thus, even weakly compressible turbulence will be dominated by shocks and shock interactions.  This suggests that compressible turbulence should have a energy exponent closer to $-2$, rather than the $-5/3$ predicted for incompressible turbulence.  Indeed, this is what is seen in the high-resolution numerical simulations of Kritsuk et al.~\cite{Kritsuk/etal:2007}.

Both the velocity and magnetic fluctuations in solar winds also follow a Kol\-mo\-go\-rov-like $-5/3$ turbulence scaling, despite the solar wind also being a conducting, compressible fluid~\cite{Roberts:2010}.  This observation has usually been explained using a variation of Sridhar \& Goldreich's~\cite{Sridhar/Goldreich:1994} theory, although Borovsky~\cite{Borovsky:2010} has cautioned that the $-5/3$ scaling could simply be due to discontinuities in the data, and may not actually be a signature of turbulence.  We will explore the possibility that discontinuities (or other singularities), produced by shocks, might be the origin of the observed velocity and density scalings.  

In summary, astrophysical fluctuations of density and velocity appear to have some of the characteristics of fully-developed turbulence, although it is difficult to reconcile this interpretation with the actual physical properties of the ISM.  In the following we review the results of computer simulations addressing these questions.  

\section{Turbulence and fragmentation --- simulations}
Computer simulations have been the tool of choice in studies of turbulent flow and have provided 
some of the most important new insights into the link between shocks, supersonic turbulence, 
and star formation in molecular clouds (eg. \cite{Vazquez-Semadeni/etal:1995, Padoan/etal:2001, Bonnell/etal:2003}. The role of ``turbulence'' is central in this picture since it 
drives the formation of filamentary structure, produces the spectrum of dense cores, and can even account
for the wide range of spins associated with these cores.  While many simulations of structure formation drive, or impose an initial velocity spectrum on the gas, the most basic question about astrophysical turbulence concerns the processes that actually drive it.  In this section, we review the insights that 
simulations have given us on these processes over this wide range of scales.

On the scales of 10~kpc characterizing a disk galaxy,  simulations of the effect of global spiral waves driven by a perturbing spiral potential have shown that the shock waves associated with these waves compress the gas into dense clouds
with an internal ``turbulent" velocity dispersion that can be associated with molecular clouds
\cite{Bonnell/etal:2006, Dobbs/etal:2006}.   The diffuse ISM
is also strongly influenced by the shocks and the PDF of this component is 
well described by log-normal density PDF \cite{Wada/Norman:2001,Wada/Norman:2007,Tasker/Bryan:2008,Tasker:2011}.   The large scale shocks
induce molecular cloud formation and create large velocity dispersion --- i.e. turbulence.   Simulations of the formation of GMCs have also idealized the process as the collision of supersonic streams of gas \cite{Vazquez/etal:2007}.  This work shows that 
at that time when collision-induced turbulence is dominated by gravitational energy of the building clouds, clouds go into global gravitational collapse.  As in combustion and pollution dispersion, turbulence is important in these processes for its extremely efficient mixing and large range of active length scales.

We have already noted that massive stars produced within sufficiently massive star clusters in 
GMCs provide strong feedback on their natal clouds and galactic environments.  One mechanism that
has been extensively simulated is the effect of multiple supernova explosions on interstellar gas.  Massive stars drive
three highly energetic phenomena into the surrounding gas:  ionization fronts originating from the intense radiation fields of massive stars, a strong stellar wind, and in the end-phase of their evolution, a supernova explosion.  Simulations of
supernova explosions in the galactic disk
show the highly textured medium and extreme density contrasts over six orders of magnitude in volume density ($10^{-4} < n < 10^2$ cm$^{-3}$) that arise from supernovae going off in a galactic disk \cite{Avillez/Breitschwerdt:2004}.    

As we descend to the scale of a molecular cloud, many studies have imposed driven or initial ``turbulent" velocity fields on gas within a periodic box setup and for which the source of the turbulence is not specified.  The earliest results of such experiments (eg. \cite{Klessen:2001}) showed that log-normal density distributions were produced independent of whether the turbulence was driven at some wavenumber $k$ or allowed to damp away (eg. review \cite{MacLow/Klessen:2004}).   As an example, Tilley \& Pudritz \cite{Tilley/Pudritz:2004} studied the combined effects of turbulence and self-gravity in a periodic box simulation in which an initial velocity field (initially set to match a Kolomogorov or Burger's spectrum) was allowed to decay in a hydrodynamic simulation.  By identifying regions that could be thought of as ``cloud cores", it was found that the mass function for such cores obeyed a log-normal distribution with a power-law tail at high masses.  It is important to note that a wide variety of density fluctuations with core-like properties were identified in these simulations.  Applying the virial theorem to these density fluctuations, it was found that a wide range of objects exist, a few of which are in equilibrium, but the bulk being either states that were bound or unbound.  The surface pressure upon density fluctuations is an important contribution towards creating gravitationally bound cores.  A large fraction of the lower mass fluctuations tended to be gravitationally unbound so that turbulence, rather than gravity, appears to dominate the dynamics of the lower mass fluctuations.    

Recent simulation show that turbulence plus gravity can reproduce the observation of power-law tails in the column density PDFs of molecular clouds~\cite{Kritsuk/etal:2011}.  The initial conditions in this periodic box simulation were prepared by applying an initial large scale random force to the gas which, after $4.8t_{\textrm{ff}}$ free-fall times, was switched off while gravity and adaptive mesh refinement (AMR) of the grid was turned on.  The adaptive refinement of the grid is necessary to follow the evolution of gravitating gas,  because it  is essential to sufficiently resolve (by at least 4 grid cells) the local gravitational instability scale (the so-called Jeans length) sufficiently so that spurious mesh-induced fragmentation of the gas does not occur \cite{Truelove/etal:1997}.  The experiments showed that an initial log-normal state was produced and that, with the onset of gravity, this log-normal distribution developed a power-law tail that extended over more than three orders of magnitude. The power-law slope for the tail of the column density was in the range $[-2.8, 2]$.  

\begin{figure}
\[\includegraphics[width=0.8\textwidth,clip=true]{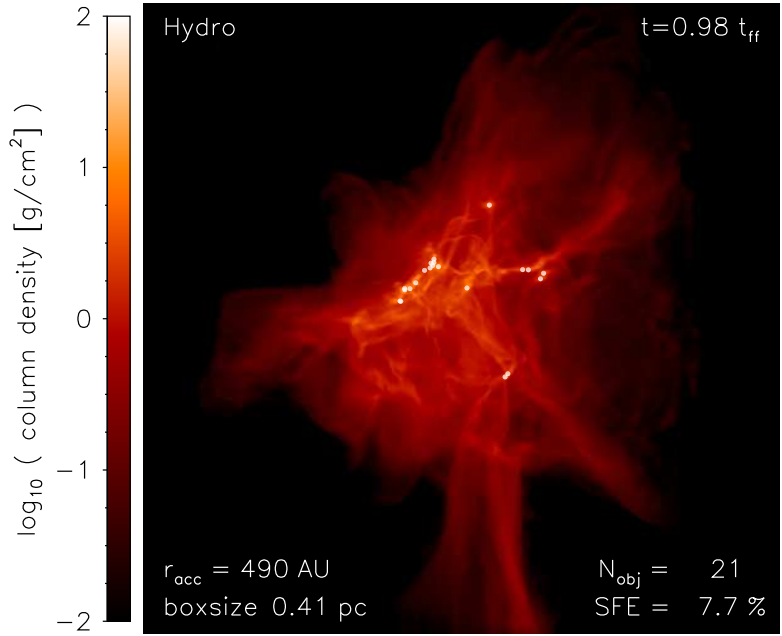}\]
\caption{The formation of stars along filaments in a FLASH AMR simulation of a cluster 
forming clump with 100 $M_{\odot}$, about 100 Jeans masses, and an initial scale of 0.4~pc (from \cite{Duffin/etal:2011}).
\label{fig:cluster}}
\end{figure}

Figure~\ref{fig:cluster} shows a snap shot of a simulation \cite{Duffin/etal:2011} of the formation of a small cluster within an initial 100 $M_{\odot}$ mass, star forming clump of gas within a molecular cloud, containing about 100 initial Jeans masses, a gas temperature 
of 10K and size of simulated region is 0.4 pc.  Collapsing regions that give rise to stars are traced by sink particles.  The simulation clearly shows that star formation is tightly associated with the filaments.  This evidence suggests that it is gravitational instability along the filament that initiates the formation of the dense collapsing regions. 

The formation of filaments in many of these simulations involves the oblique collisions of planar shock fronts that arise from the initial velocity field.  The collision of two sheets (shock wave-fronts) is a line.  Thus filaments represent the densest structures that supersonic turbulence, or, in our view, a collection of shock waves, can produce.  It is therefore not surprising that dense cores would appear to form along filaments, as many simulations have shown.   The formation of a filament by such a collision process is  clearly seen in work by Banerjee et al~\cite{Banerjee/etal:2006} where the collision of two quasi-planar shocks is seen to give rise to a density filament. 

Another key aspect of the ``turbulent fragmentation'' picture of structure formation is that it provides a natural explanation for the origin of angular momentum for cores, and hence for the origin of protostellar disks.  Rotational energies are generally a small fraction of the gravitational energy of 
the system: the average value of the ratio of the rotational to gravitational energy is $\beta_{rot} \simeq 0.01$, but its value spans several orders of magnitude: for both low mass with a range $2\times 10^{-3} - 1.4$ \cite{Goodman/etal:1993} as well as high mass cores \cite{Pirogov/etal:2003} in the range $4 \times 10^{-4} - 7 \times 10^{-2}$.   While rotational energy is thus too low to play a significant role in the dynamics of the cores, it is significant in 
establishing the initial conditions for the formation of protostellar disks on much smaller scales (a few to several hundred AU) Angular momentum can arise as a consequence of the oblique collision of shock waves~\cite{Jappsen/etal:2004,Tilley/Pudritz:2004}.  The range of spins of such cores can be quite broad, extending over several orders of magnitude.   In simulations by Banerjee et al~\cite{Banerjee/etal:2006} the angular momentum of this gas arises from the fact that the filament is clearly associated with the intersection of two sheets, which are presumably the shock fronts that generated from the initial supersonic velocity field set up for the cloud.  The large scale flow along the filament has sufficient angular momentum that an accretion disk is built up.

We now turn to discuss recent results~\cite{Kevlahan/Pudritz:2009} in which we showed that Kolmogorov scaling over many length scales and the observed skewed log-normal density PDF could arise from repeated shock interactions, without the need for fully-developed turbulence.   In addition, we  discuss the physical origin of the angular momentum (i.e. vorticity) that is generated in interacting shocks, as well as in curved shocks.

\section{Shock forcing of turbulence}
We have noted that the collision of shock waves generates filaments, and that the simulations typically and rapidly produce lognormal structure in the gas.   In this section, we show that  
compressible turbulence fluctuations produce lognormal part of the distribution through interaction of eddy shocklets.  Thus, lognormal structure which is evident in the IMF, has its roots in the basic dynamics of interacting shock waves.  While gravity certainly can produce power-law tails observed in molecular cloud density PDFs - we 
show that feedback processes produce spherical shocks which also give rise to the power law tails.  The role of spherical shocks (or spherical expansion waves) has been confirmed in a recent numerical study~\cite{Kritsuk/etal:2011}.  

\subsection{Regular intersection of planar shocks}
\begin{figure}
\[\includegraphics[width=\textwidth,clip=true]{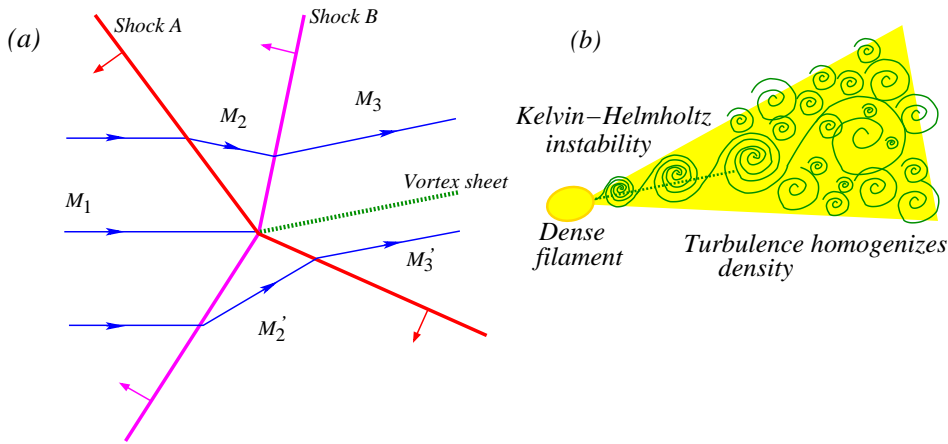}\]
\caption{(a)~Regular intersection of two plane shocks of unequal strength (the figure show a two-dimensional section).  $M_1$, $M_2$, $M_2'$, $M_3$ and $M_3'$ are the Mach numbers of the flow in the various regions. Note that the shocks are slightly bent, the streamlines are rotated and a vortex sheet is generated downstream of the shocks.  (b)~Instability of the vortex sheet produces turbulence that mixes and homogenizes the density away from the shock intersection point.
\label{fig:shocks}}
\end{figure}
The simplest type of shock interaction is the regular intersection of two plane shocks of unequal strength~\cite{Liepmann/Roshko:1957,Chapman:2000}.  Figure~\ref{fig:shocks}(a) shows the basic features of such an interaction: both shocks are bent slightly, the flow is rotated and a vortex sheet (or slip stream) is generated downstream of the shock.  As sketched in Figure~\ref{fig:shocks}(b), the vortex sheet is unstable and produces a sequence of vortices via the Kelvin--Helmholtz instability. These vortices in turn become unstable and generate three-dimensional turbulence via a secondary instability (note that the intersection of two plane shocks of equal strength will 
not generate a vortex sheet).  As seen in the simulations described in the previous section,  a dense filament is observed just behind the shock intersection point.  One expects that the density further downstream of the shock should be much more homogeneous due to the efficient turbulent mixing produced by the unstable vortex sheet.  Note that two shocks generically intersect in a line, and this explains the fact that the most over dense regions have a filamentary structure.

It is important to note that, as discussed below, a vortex sheet instantaneously distributes kinetic energy over all length scales and has a power law energy spectrum $E(k)\sim k^{-2}$.  This initial spectrum becomes shallower as the vortex sheet becomes unstable.

In summary, the regular intersection of two unequal planar shocks generates both a dense filamentary structure and turbulence (via the destabilization of the vortex sheet produced behind the shock intersection line).  In the following section, we review how multiple shock passages may build up the lognormal density distribution and $k^{-5/3}$ energy spectrum observed in the ISM.  These multiple shock passages may involve the intersecting plane shocks just discussed, or individual self-focusing shocks.

\subsection{A shock mechanism for the density PDF and energy spectrum}
In this section we review the shock-based mechanisms proposed in recent work by Kevlahan \& Pudritz~\cite{Kevlahan/Pudritz:2009} to explain the lognormal PDF of density and turbulence-like $k^{-5/3}$ spectrum of kinetic energy observed in the ISM.  We focus here on the role of individual curved shocks, but the intersecting plane shocks discussed in the previous section play a similar role.

Kevlahan~\cite{Kevlahan:1997} derived an expression for the vorticity jump across a shock wave that includes the effect of the upstream turbulence.  If we assume that the turbulence is isentropic and quasi-steady (with respect to the passage time of the shock), then the
vorticity jump in the binormal direction $\bm{b}$
\beq
\delta\bm{\omega}\cdot\bm{b} = \frac{\mu^2}{1+\mu} \frac{\partial M_s}{\partial S}
+ \frac{1}{M_s}\left(\frac{\mu}{1+\mu} M_s^2-1\right) \left[\frac{\partial\frac{1}{2}M_t^2}{\partial S}
+ \bm{\omega}\times\bm{u}\cdot\bm{s}\right] + \mu\bm{\omega}\cdot\bm{b},
\label{eq:voder7}
\eeq
where $\mu$ is the normalized density jump (i.e. the shock strength), $M_s$ is the Mach number of the shock, $\partial/\partial S$ is the tangential part of the directional derivative, $M_t$ is the turbulent Mach number of the upstream flow and $\bm{u}$ and $\bm{\omega}$ are the velocity and vorticity of the upstream turbulence normalized by sound speed.  Equation
(\ref{eq:voder7}) may be simplified for strong shocks by using
the approximation $\mu\approx 2/(\gamma-1)$.  The density jump is given by 
\beq
{\delta\rho \over\rho} \equiv \mu = \frac{M_s^2-1}{1+1/2(\gamma-1)M_s^2}.
\label{eq:density}
\eeq

The first term on the right hand side of (\ref{eq:voder7}) represents
vorticity generation due to the variation of the shock speed $M_s$
along the shock -- e.g. due to shock curvature -- by symmetry it is exactly zero for spherical and
cylindrical shocks.  The second term is baroclinic generation of vorticity due to misalignment of pressure and density gradients across the shock and the final term the jump in vorticity due to the conservation of angular momentum across the shock.

Because shocks are nonlinear waves (unlike acoustic waves), $M_s$ is larger in regions of concave curvature and
smaller in regions of convex curvature (with respect to the
propagation direction of the shock).  This difference in shock
strength increases over time and eventually causes curved shocks to
{\em focus\/} at regions of minimum curvature, developing a flat shock disk
bounded by regions of very high curvature (often called
kinks)~\citep{Kevlahan:1996}.  This structure is called a Mach disk, and generates vortex sheets downstream of the shock as in the case of intersecting plane shocks. The discontinuous
shock strength at the kinks is called a shock--shock~\cite{Whitham:1974}, and is essentially the same as the intersection line of a pair of intersecting plane shocks. 

In the ISM focusing of individual shocks could arise due to reflection off density gradients
(e.g. vertically stratified structure in disks), density inhomogeneities, or small variations in shock curvature in blast waves.

Since the first term on the right hand side of
(\ref{eq:voder7}) is approximately singular at the location of a kink,
vortex sheets develop in the flow downstream
of the kinks.  These vortex sheets
themselves have an energy spectrum $E(k)\sim k^{-2}$, and generate
turbulence exponentially fast via the Kelvin--Helmholtz instability.
Note that this produces a $-2$ spectrum in the downstream
flow, even when the shock is no longer present.  In other words, the $-2$ energy spectrum scaling could be the {\em trace\/} of a shock, rather than requiring the presence of a shock.

Kevlahan \& Pudritz~\cite{Kevlahan/Pudritz:2009} showed for the first time how multiple passages of focused shocks could produce an energy spectrum shallower than $-2$.
Previously, the relevance of multiple shock passages for astrophysical flows was
established by \cite{Kornreich/Scalo:2000}, who found that the average
time between shock passages in the ISM is ``small enough that the
shock pump is capable of sustaining supersonic motions against
readjustment and dissipation.''

The energy spectrum of the three-dimensional flow 
after one, two and three passages of a focused
shock with principal wavenumbers $k_1=k_2=1$, $M_s=6$ and zero velocity initial condition was analyzed by \cite{Kevlahan/Pudritz:2009} . The key result is that the initial $k^{-2}$ scaling of the energy spectrum (due to
the velocity discontinuity associated with the vortex sheet downstream
of the shock--shocks) becomes gradually shallower with each shock
passage.  This redistribution of energy to smaller scales is due to
the quadratic baroclinic terms depending on the (inhomogeneous)
upstream flow in the vorticity jump equation (\ref{eq:voder7}).
Although the effect is entirely kinematic, these quadratic terms
redistribute energy amongst scales in a way analogous to the quadratic
nonlinearity in the Navier--Stokes equations.
After three shock passages the energy spectrum has a scaling similar
to $k^{-5/3}$.  More shock passages would produce an even shallower
power-law.  Moreover,  the form of the
energy spectrum is relatively insensitive to the choice of initial
condition.

Simulations have suggested previously that the structure observed in the interstellar medium and molecular
clouds probably derives from shock-driven processes.  The results presented above from Kevlahan \& Pudritz~\cite{Kevlahan/Pudritz:2009} give an alternative explanation of the energy spectrum scalings close to $-5/3$ observed in astrophysical fluctuations.  They also suggest why the spectrum is shallower than the scaling $-2$ expected for compressible turbulence (due to the eddy shocklets generated spontaneously in strongly compressible flows).  The observed $E(k)\approx k^{-5/3}$ ``big power-law in the sky'' may be due to primarily to kinematic effects associated with repeated shock passages, and may not in fact require fully-developed turbulence.  This could also explain why the power-law extends over such a huge range of length scales (spanning several
different physical regimes, from the diffuse ISM to molecular gas),
since the power-law of a shock-driven flow is due to a singularity,
and so extends over all scales (until the viscous cut-off).  

We have shown that interacting shocks generate a log-normal distribution of density.  As mentioned above, compressible turbulent flows with $M_t>0.3$
spontaneously generate small, relatively weak and highly curved
shocks, called `eddy shocklets'~\cite{Kida/Orszag:1990}.  It is
therefore likely that a particular region of space is hit
multiple times by eddy shocklets of varying strengths.  If we now assume that
the density doesn't change significantly between shock passages
(i.e. that the density changes are primarily due to shock compression),
then from equation (\ref{eq:density}) the density after $n$ shock
passages is
\beq
\rho^{(n)}(x) = \prod_{j=0}^n (1+\mu^{(j)}(x))
\eeq
where the density is normalized in terms of the initial uniform density
$\rho_0$.  We now treat the shock strengths $\mu^{(j)}(x)$ as
identically distributed random variables.  Then, since
$(1+\mu^{(j)}(x))>0$ we can take the logarithm of both sides and apply
the central limit theorem to the resulting sum.  This shows that the
logarithm of density is normally distributed, i.e. the density PDF
follows a log-normal distribution,
\beq
P(\rho) =
\frac{1}{\sqrt{2\pi}\sigma\rho}\exp\left(-\frac{(\log(\rho)-\overline{\log\rho})^2}{2\sigma^2}\right).
\label{eq:lognormal}
\eeq
Note that application of the central limit theorem to derive
(\ref{eq:lognormal}) requires only that the random variables
$\log(1+\mu^{(j)}(x))$ have finite mean and variance~\cite{Vazquez-Semadeni:1994}.

\begin{figure}
\[\includegraphics[width=0.8\textwidth,clip=true]{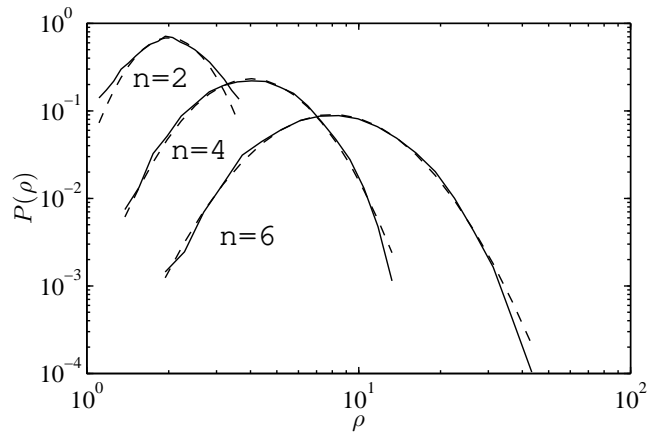}\]
\caption{Convergence to a log-normal PDF of density after $n=2,4,6$
shock passages.  - - -, log-normal distribution, --- PDF of density (adapted from \cite{Kevlahan/Pudritz:2009}).
\label{fig:lognormal}}
\end{figure}

Kevlahan \& Pudritz~\cite{Kevlahan/Pudritz:2009} pointed out that the convergence to the lognormal distribution very rapid.   Figure~\ref{fig:lognormal} (reproduced from \cite{Kevlahan/Pudritz:2009} ) shows that as few as three or four shock passages generates a very good approximation to the log-normal distribution.
This explanation is consistent with observations from experiments that the lognormal distribution of density develops very quickly, and seems to be associated with the first shock interactions~\cite{Nordlund/Padoan:1999}.  Thus, the log-normal distribution should develop well before self-gravity becomes important.  In fact, as proposed below, the denser cores associated with the log-normal distribution are likely the catalysts for the self-gravity driven collapse that produces the power-law tail at high densities.

\subsection{Feedback effects and power-law tails}

As first suggested in ~\cite{Kevlahan/Pudritz:2009}, a power-law tail of the density distribution can be produced by spherical shock waves which interact with a pre-existing lognormal distribution.  These shocks could be due to blast waves or stellar wind (once the first stars have formed), or could take the form of an expansion wave generated by a rapidly collapsing core (as suggested by Kritsuk et al~\cite{Kritsuk/etal:2011}).  The origin of the shock wave will change the power law scaling, but it is the spherical symmetry of the shock that produces a power-law scaling at large densities.  

Following \cite{Kevlahan/Pudritz:2009}, we assume a blast wave  solution for strong spherical shocks.  Dokuchaev~\cite{Dokuchaev:2002} generalized the Sedov--Taylor self-similar blast wave solution to the sustained energy injection case $E(t)\propto t^p$.  $p=0$  corresponds to an instant shock (e.g. such as a super nova),  while $p=1$ corresponds to continuous energy input (i.e. a stellar wind).  

As in the previous case, we take a probabilistic view and assume that the PDF of finding a particular value of gas density $\rho_1$ is proportional to
the the space--time volume where the density exceeds $\rho_1$,
\beq
P(\rho>\rho_1) \propto \int_0^{t(\rho_1)} R^3(t) \dee{t}
\eeq
where $R(t)\propto t^{(2+p)/5}$ is the radius of the spherical shock
at time $t$ and $t(\rho_1)$ is the time at which the density behind
the shock is equal to $\rho_1$.  Using the relation $M_s(t)\propto
R(t)/t$, equation (\ref{eq:density}) can be inverted to find
$t(\rho_1)\propto \rho^{5/(2(-3+p))}$.

Using the definition of the PDF, we find that the PDF of density due to
the interaction of a homogeneous gas with a spherical blast wave has
the form
\beq
P(\rho) = \frac{d}{d \rho}\int_0^{t(\rho)} R^3(t) \dee{t}
\propto \rho^{-(17+p)/(6-2p)}
\label{eq:pdf_sphere}
\eeq
(where we have re-labelled $\rho_1$ as $\rho$). Therefore, the density
PDF is a power-law with exponent $-17/6\approx -2.8$ for an instant
shock and $-9/2 = -4.5$ for an injection shock.  These slopes are
significantly steeper than the Salpeter value for the mass PDF of
$-1.35$.  However, the actual relation between the density PDF and the
mass PDF depends on the precise assumptions made about the scaling of
clumps (i.e. mass equals density times a lengthscale cubed).  Thus,
one should not necessarily expect the same index for both the density
and mass PDFs (although the power-law form should be robust). Furthermore, as mentioned earlier, analysis of dust extinction maps~\cite{Kainulainen/etal:2009} gave scalings for the power-law tail of the density PDF of -1.5 to -4.2, roughly consistent with the spherical shock model.

It is important to note that observations measure PDFs of {\em column density\/} (i.e. density integrated along a line of sight), rather the PDFs of density itself.  If there are many randomly distributed spherical shocks these two measures should be similar.  However, if there is only one infinitely large spherical shock the resulting density would have to be integrated to give a column density as in~\cite{Kritsuk/etal:2011}. If the density is given by $\rho\sim r^{-n}$ then the column density is 
\beq
2\int_0^\infty \rho(\sqrt{R^2+x^2}) dx \propto R^{1-n},
\eeq
which also has a power-law PDF, but with slope $-2/(n-1)$.  In the blast wave model $\rho\sim r^{-2(3-p)/(2+p)}$ and thus $n=3$ for a blast wave and $n=4/3$ for an injection shock.  Thus, if we assume a single blast wave the column density would scale like $-1$, while for the injection shock the column density would scale like $-6$.  Kritsuk eta al.~\cite{Kritsuk/etal:2011} found column density scalings for self-similar gravitational collapse of $-2,-2.8, and -4$, depending on the model.  It therefore seems more likely that the observed power-law is due either to expansion shock waves produced by self-gravitational collapse, or to a collection of spherical blast waves.  However, as noted below, the probabilistic interpretation helps explain the hybrid form of the PDF (both log-normal and power-law).

\begin{figure}
\[\includegraphics[width=0.8\textwidth,clip=true]{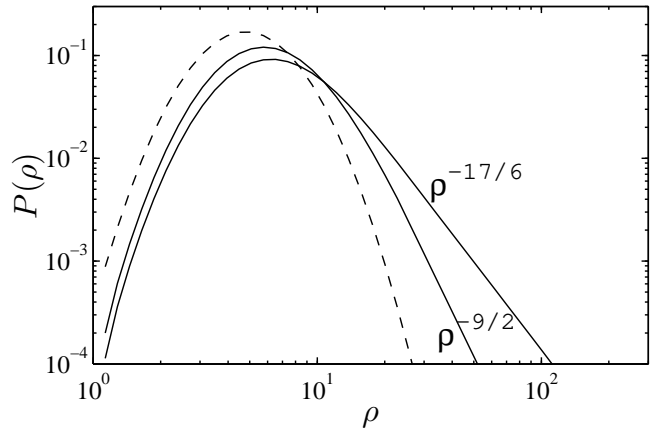}\]
\caption{Generation of a power-law PDF at large densities in a nearly
isothermal gas by a spherical blast wave interacting with a
log-normally distributed density field. - - - is the initial
log-normal PDF and the slopes -17/6 and -9/2 correspond to an instant
shock and an injection shock respectively.  Note that the upper limit
of the power-law range is proportional to $1/(\gamma-1)$, and thus we
expect the largest power-law ranges for nearly isothermal gases,
i.e. those with $\gamma\approx 1$. (Adapted from \cite{Kevlahan/Pudritz:2009})
\label{fig:skewed}}
\end{figure}

Finally, we can find the PDF of density resulting from the interaction of a
spherical shock with a log-normally distributed density field as the convolution of the PDF (\ref{eq:pdf_sphere}) with the
log-normal distribution of density (\ref{eq:lognormal}).  This
produces a PDF which is log-normal for small densities, and has a
power-law $\rho^{-(17+p)/(6-2p)}$ for high densities up to a maximum
density proportional to $1/(\gamma-1)$.  Figure~\ref{fig:skewed}, reproduced from \cite{Kevlahan/Pudritz:2009} shows the resulting hybrid PDF.  If no more spherical shocks are generated the distribution will slowly revert to a simple log-normal form.  Thus, one expects that a power-law tail is a signature of a star-producing region, where shocks are produced either by intense stellar winds or by spherical expansion waves associated with self-gravitational collapse.  The remaining question is whether such shock waves are able to account for so many decades of gas compression
as can gravity.

\section{Conclusions}
The structure of interstellar gas in general, and star-forming molecular clouds in particular, is shaped to 
a very important degree by interacting shock waves.  Using two basic shock models (intersecting plane shocks and self-focusing curved shocks) we have shown that this provides a more
self-consistent picture of the ISM and the conditions leading to star formation than traditional ideas
of ``supersonic turbulence".  The log-normal aspect of the gas density is imprinted on the structure
of the IMF of stars.  The power-law tail at high masses can be produced by several mechanisms --- certainly
by gravity --- but also by the shock waves resulting from radiative feedback processes during star formation.  
The generality of this behaviour of shock waves gives considerable support for the idea that, despite
the complexity in the detailed physics, the structure and possible universality of the form of the kinetic energy spectrum and IMF are an expression of the physics of shock waves in gravitating media.  

\begin{acknowledgements}  REP wishes to thank the organizers of the 
Turbulent Mixing and Beyond, 2010 in Trieste, for the opportunity to give a talk on this subject.  We are grateful to
Snezhana Abarzhi for her invitation,  encouragement and patience as an Editor of this special volume.   We also
thank an anonymous referee for useful remarks on our manuscipt.  
The research of both NK and REP was supported by NSERC Discovery Grants.
\end{acknowledgements}

\bibliographystyle{vancouver}
\bibliography{bib_tmb}
\end{document}